\documentclass[twocolumn,showpacs,preprintnumbers,amsmath,amssymb]{revtex4}

\usepackage{graphicx}
\usepackage{dcolumn}
\usepackage{bm}

\begin{document}


\title{Experiment towards continuous variable entanglement swapping: Highly
correlated four--partite quantum state}

\author{Oliver Gl\"ockl$^{1}$}
\email{gloeckl@kerr.physik.uni-erlangen.de}
\author{Stefan Lorenz$^1$}
\author{Christoph Marquardt$^1$}
\author{Joel Heersink$^1$}
\author{Michael Brownnutt$^1$}
\author{Christine Silberhorn$^1$}
\author{Qing Pan$^{1,2}$}
\author{Peter van Loock$^1$}
\author{Natalia Korolkova$^1$}
\author{Gerd Leuchs$^1$}

\affiliation{$^1$Zentrum f\"ur Moderne Optik, Physikalisches Institut,
Universit\"at Erlangen--N\"urnberg, Staudtstra{\ss}e 7/B2, 91058 Erlangen,
Germany\\
$^2$The State Key Laboratory of Quantum Optics and Quantum
Optics Devices, Institute of Opto--Electronics, Shanxi University, Taiyuan,
030006, P.R.China}

\date{\today}

\begin{abstract}
We present a protocol for performing entanglement swapping with intense pulsed
beams. In a first step, the generation of amplitude correlations between two
systems that have never interacted directly is demonstrated. This is verified in
direct detection with electronic modulation of the detected photocurrents. The
measured correlations are better than expected from a classical reconstruction
scheme. In an entanglement swapping process, a four--partite entangled state is
generated. We prove experimentally that the amplitudes of the four optical modes
are quantum correlated 3 dB below shot noise, which is consistent with the
presence of genuine four--party entanglement.
\end{abstract}

\maketitle

\section{Introduction}
Entanglement is the basic resource for quantum information applications. We are
dealing with intense, pulsed light which is described by continuous quantum
variables and which allows for efficient detection schemes and reliable
sources. The generation of entanglement shared by two parties is now achieved
routinely in the laboratories, e.\,g.\, entanglement of the quadrature
components of an electromagnetic field. These two party entangled states may
enhance the capability of the two parties to communicate. There are already
several experimental realisations of quantum information and quantum
communication protocols over continuous variables, exploiting entanglement, for
example, quantum dense coding \cite{LI02} and quantum teleportation
\cite{FUR98}.

Initially, quantum communication dealt almost exclusively with discrete
two--valued quantum variables. The first demonstration of teleportation,
i.\,e.\, the transfer of a quantum state from one party to another was reported
by Bouwmeester et al.\,\cite{BOU97}. The scheme was close to the theoretical
proposal by Bennett et al.\,\cite{BEN93}. In the experiment, the polarization
state of a single photon was teleported using a pair of polarization entangled
photons. However, the experiment was not an unconditional teleportation, as only
one of the four possible results of the Bell--state measurement can be
discriminated against the others. As a next step, quantum entanglement swapping,
i.\,e.\, the teleportation of entanglement, was demonstrated experimentally by
Pan et al.\,\cite{PAN98}. More recently, unconditional teleportation of an
unknown polarization state was demonstrated by Kim et al.\,\cite{KIM01a},
employing nonlinear interactions to discriminate the entire set of Bell--states,
however at low efficiency. These initial experiments relied upon the
polarization of a single photon to encode the qubit. Recently, long distance
teleportation was demonstrated by encoding the qubit into the superposition of a
single photon in two different locations (time bins) \cite{MAR03}. In all these
experiments, discrete quantum variables were teleported.

For continuous variables, such as the amplitude and the phase quadratures of an
electromagnetic field, which are used in this paper, only the
quantum teleportation of coherent states has been demonstrated so far. The first
experiment was reported by Furusawa et al.\,\cite{FUR98} and was based on a
proposal by Braunstein and Kimble \cite{BRA98}. The advantage of this experiment
is that unconditional teleportation was demonstrated, i.\,e.\, no postselection
of successful events was necessary. However, the quality of the teleportation,
i.\,e.\,the fidelity is limited by the quality of the correlations of the
auxiliary entangled beam pair. In that first experiment the fidelity was
0.58. Recently, there have been reports on further improvements in teleportation
experiments of coherent states, with fidelities of 0.62 \cite{ZHA02} and 0.64
\cite{BOW02a}.

So far, no entanglement swapping with continuous variables has been
reported. However, entanglement swapping is interesting for several reasons:
First, among all quantum states of the light field coherent states are closest
to classical states while entanglement swapping refers to the teleportation of
a highly non--classical state. Second, the success of entanglement swapping can
be checked easily by verifying the correlations generated in the entanglement
swapping process. In general, for more complex applications towards quantum
networking it is desireable to achieve entanglement swapping combined with
entanglement purification \cite{BRI98}, as it enables the distribution of
entanglement and non--classical correlations over large distances between
systems that have never interacted directly.

Multipartite entanglement, the entanglement shared by more than two parties, is
also a useful resource for quantum networking. For example, the distribution of
quantum information to several receivers, called telecloning \cite{LOO01} or for
quantum secret sharing \cite{HIL99} are based on multi--party entangled states.

In this paper, we present our work towards entanglement swapping using
intense beams. We present a possible entanglement swapping scheme and describe
and characterize our entanglement sources. For entanglement swapping, two
independent EPR--sources are needed which are then made to interfere. By a
direct analysis of the detected photocurrents we prove that strong correlations
are created between the amplitude quadratures in the entanglement swapping
process. This experiment is a first step towards the teleportation of a highly
non--classical state. We also show that with the same basic resource,
i.\,e.\, the coupling of two entanglement sources, a highly correlated
four--party state is generated. A preliminary theoretical analysis, neglecting
the excess thermal phase noise in the EPR--sources and hence assuming them to be
pure, indicates that the generated state is indeed a genuinely four--party
entangled state.

\section{Entanglement swapping protocol}

\begin{figure}
\includegraphics[scale=0.32]{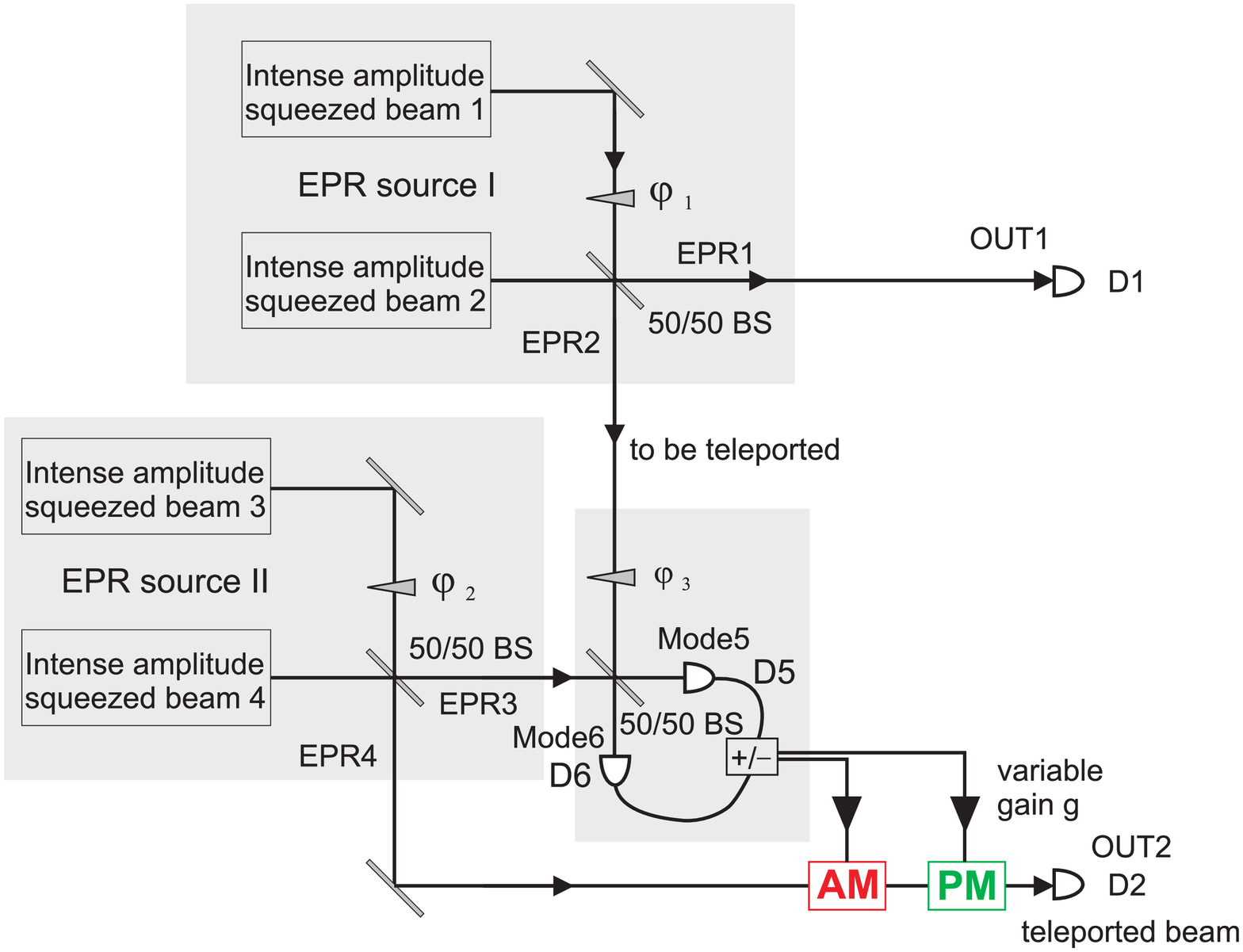}
\caption{\label{setup}Schematic drawing of the entanglement swapping experiment}
\end{figure}

The scheme for entanglement swapping is outlined in Fig.~\ref{setup}. Two
pairs of quadrature entangled beams denoted EPRI and EPRII are generated by
linear interference of two amplitude squeezed beams. Initially, the beam
pair labelled EPR1 and EPR2 and the one labelled EPR3 and EPR4 are two
independent entangled pairs. The experiment aims at achieving entanglement
between the beams 1 and 4. This requires the teleportation of beam EPR2 to the
output mode EPR4 and is referred to as entanglement swapping. For that purpose,
one beam from each entanglement source, i.\,e.\, EPR--beam 2 and EPR--beam 3
are combined at a 50/50 beam splitter with the phase $\varphi_3$ adjusted such
that the two output beams are equally intense (Fig.~\ref{setup}). The
resulting beams are denoted Mode5 and Mode6.

The Bell--state measurement is completed by the joint detection of Mode5 and
Mode6 (see Fig.~\ref{setup}). The signals obtained from that
measurement are used to modulate the amplitude and the phase of EPR4 to yield
OUT2, which in the ideal case should now be a copy of the mode EPR2 and show
strong non--classical correlations with EPR1, now called OUT1. The signals
needed for the modulation are described by $(\delta\hat{X}_5+\delta\hat{X}_6)$
and $(\delta\hat{Y}_5-\delta\hat{Y}_6)$, where $\delta \hat{X}$ and $\delta
\hat{Y}$ denote the fluctuations in the amplitude and the phase quadrature. The
quadrature components \cite{WAL94} of the electromagnetic field are defined by
$\hat{X}=\hat{a}^{\dagger}+\hat{a}$ and $\hat{Y}=i(\hat{a}^{\dagger}-\hat{a})$.
The sum-- and difference variances can be obtained in direct detection without
local oscillator techniques, provided the measurement is performed on intense
beams \cite{KOR00b, LEU99, ZHA00}. In a next step, the fluctuating results of
the photon number measurements on Mode5 and Mode6, $\delta\hat{n}_5$ and
$\delta\hat{n}_6$ are detected. Taking simultaneously the sum and the difference
of the corresponding photocurrents, that is $\delta\hat{n}_5+\delta\hat{n}_6$
and $\delta\hat{n}_5-\delta\hat{n}_6$, signals are obtained which are
proportional to the sum and difference quadratures mentioned above. The
corresponding photocurrents which are transmitted over the classical channel to
the modulators are denoted $i_{\rm Bell}^{+}=i5+i6$ and $i_{\rm
Bell}^{-}=i5-i6$, respectively.

An optimum gain $g$ for the modulation can be chosen such that after the
entanglement swapping process the highest possible correlations between the
output beams OUT1 and OUT2 are generated. The gain $g$ describes to what
degree initial fluctuations of one beam are transferred onto an output
mode after detection of the initial mode and subsequent modulation of the output
mode with that signal. In the case of infinite input squeezing $g=1$ is
optimal, while for finite squeezing values of $g<1$ are better. The value of
$g$ also depends on the degree of excess noise in the anti--squeezed quadrature,
i.\,e.\, the optimum value for $g$ is different for non--minimum
uncertainty squeezed states and for minimum uncertainty states. The
optimum value for $g$ has to be chosen closer to one the more excess phase noise
is present.

It was shown \cite{LOO99} that for any finitely squeezed minimum uncertainty
vacuum states, EPR--correlations between the output modes OUT1 and OUT2
can always be generated independent of the degree of input squeezing using an
optimized gain. This also holds true for intense squeezed beams that we
employ in our experiment. However, for states with high excess noise in the
phase quadrature, say about 20dB as in our experiment, the optimum gain is close
to $g=1$. Hence there is a 3dB penalty in the correlations created between the
output states. Therefore, in this case, the generation of nonclassical
EPR--correlations in the entanglement swapping process  requires more than 3dB
initial squeezing.

\section{Experimental setup and measurement results}
\subsection{Entanglement source}

\begin{figure}
\includegraphics[scale=0.34]{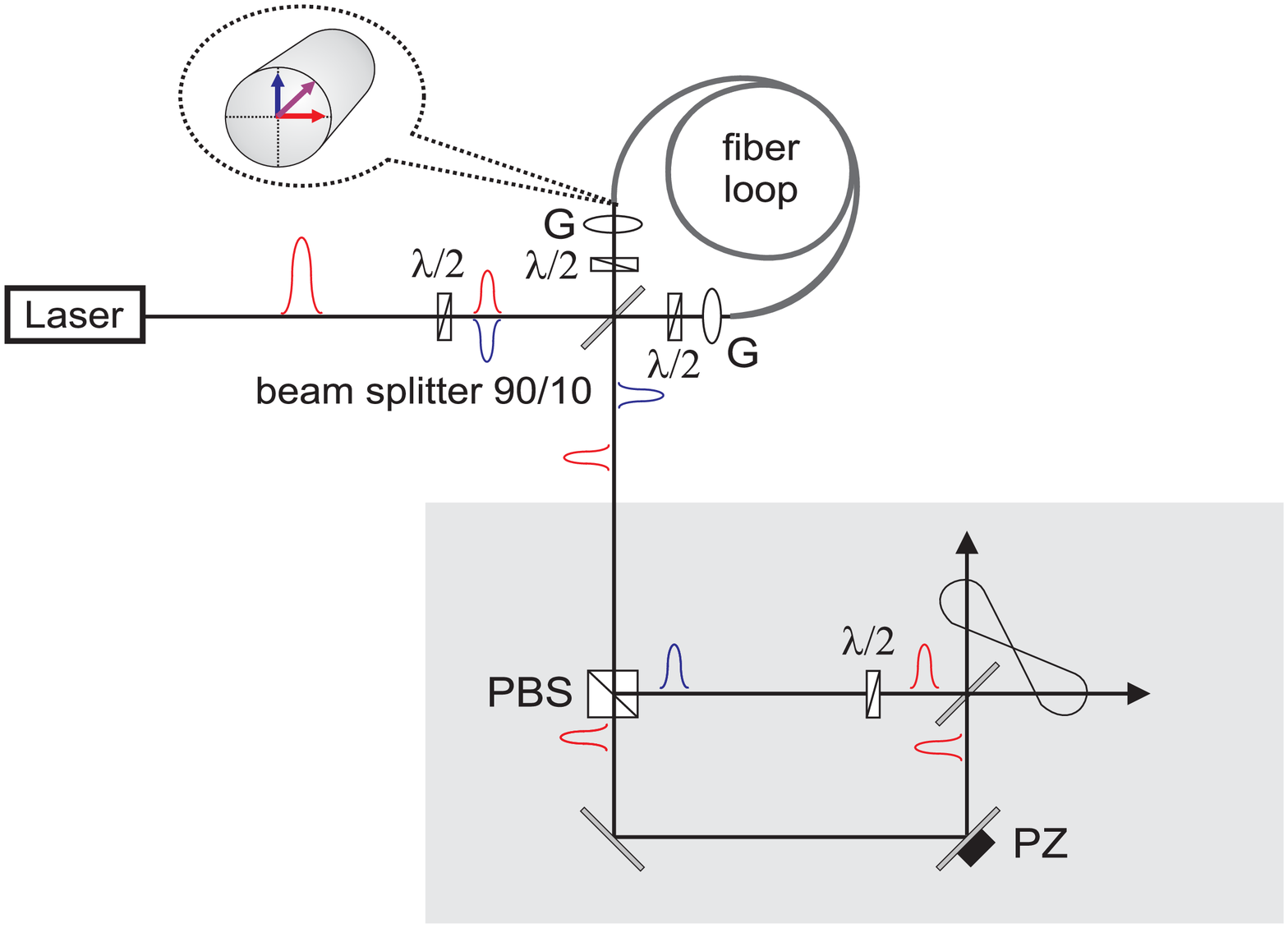}
\caption{\label{squeezing}Generation of squeezing using an asymmetric fiber
Sagnac interferometer. $\lambda/2$: half wave plate, G: gradient index lens,
PBS: polarizing beam splitter, PZ: piezo actuator. Two independently squeezed
modes are generated as the light propagates on both optical axes of the optical
fiber. The gray shaded area shows the interference of both amplitude squeezed
fields to generate entanglement.}
\end{figure}

The intense squeezed beams are generated using an asymmetric fiber Sagnac
interferometer, exploiting the Kerr nonlinearity. The setup is
depicted in Fig.~\ref{squeezing}.

In the asymmetric fiber Sagnac interferometer, two counter propagating pulses,
one strong pulse and one weak pulse, are coupled into a polarization maintaining
fiber. Due to the Kerr nonlinearity, the strong pulse
acquires an intensity dependent phase shift, while the weak pulse is largely
unaffected by nonlinear effects. Thus, the initially circular shaped uncertainty
area in phase space is transformed into an ellipse. By interference with the
weak pulse, the uncertainty ellipse is reoriented in phase space, resulting
in direct detectable amplitude squeezing \cite{SCHM98}. Choosing an input
polarization of about $45^{\circ}$ with respect to the optical axes of the
fiber, two independently squeezed beams in s-- and in p--polarization can be
generated simultaneously\cite{KOR00,SIL01}. In the following this
interferometer is referred to as a double squeezer.

Quadrature entanglement is generated by linear interference of two amplitude
squeezed fields with proper phase relation \cite{LEU99,SIL01}. Due to the
birefringence of the fiber, the relative delay of the pulses has to be
compensated before they are combined on a 50/50 beam splitter to generate
entanglement (see the shaded region in Fig.~\ref{squeezing}). The
EPR--entanglement is maximized when the interference phase is such that the two
output beams have equal optical power. An active feedback control to stabilize
the interference phase is used.

The entanglement is characterized in terms of the non--separability criterion
for two mode Gaussian states \cite{DUA00,SIM00}. The non--separability criterion
can be expressed in terms of observable quantities, the so called squeezing
variances \cite{KOR02a}:
\begin{eqnarray}
V_{\rm sq}^{\pm}(\delta \hat{X})&=&\frac{V(\delta\hat{X}_{\rm
i}\pm g\delta\hat{X}_{\rm j})} {V(\delta\hat{X}_{\rm i,
coh}+ g\delta\hat{X}_{\rm j, coh})},\\
V_{\rm sq}^{\mp}(\delta \hat{Y})&=&\frac{V(\delta\hat{Y}_{\rm
i}\mp g\delta\hat{Y}_{\rm j})} {V(\delta\hat{Y}_{\rm i,
coh}+ g\delta\hat{Y}_{\rm j, coh})}
\end{eqnarray}
where $V(\hat{A})=\langle \hat{A}^2 \rangle- \langle \hat{A} \rangle^2$ denotes
the variance of an operator $\hat{A}$ and $i\neq j$ and $g$ is a variable gain.
The quadrature components labelled with index  ``coh'' are those of a coherent
state. The non--separability criterion then reads \cite{KOR02a}:
\begin{equation}\label{peres}
V_{\rm sq}^{\pm}(\delta \hat{X})+V_{\rm sq}^{\mp}(\delta \hat{Y})<2.
\end{equation}

\subsection{Towards entanglement swapping---direct analysis of the photocurrent}

\begin{figure}
\includegraphics[scale=0.4]{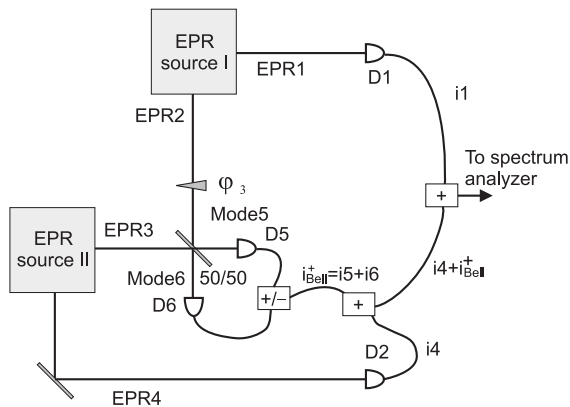}
\caption{\label{entswap}Entanglement swapping, direct analysis of the
photocurrents}
\end{figure}

In this section we describe a scheme that permits us to check for correlations
in the amplitude quadrature created by the entanglement swapping process by
direct analysis of the photocurrents. The signal from the Bell--measurement,
$i_{\rm Bell}^{+}$, that will be used for the modulation is a classical signal.
If the emerging entangled pair, OUT1 and OUT2, were to be used as a quantum
resource, the signal from the Bell--state measurement has to be used to
modulate the optical mode EPR4 (see Fig.~\ref{setup}). However, if the
mode OUT2 is just detected to verify the success of entanglement swapping,
modulation of EPR4 can be substituted by direct summation of the photocurrents.
Thus, the measured photocurrents are identical in the case of optical modulation
of EPR4 and detecting OUT2 and in the case of measuring EPR4 and adding the
photocurrent $i_{\rm Bell}^{+}$, giving $i4 + i_{\rm Bell}^{+}$ (see
Fig.~\ref{entswap}). In this context, we performed the following experiment
towards entanglement swapping as it is depicted in Fig.~\ref{entswap}.

Two entanglement sources were set up, each consisted of a double squeezer and a
subsequent interferometer to generate entanglement. In each squeezer, 8m of
polarization maintaining fiber (FS--PM--7811 from 3M) was used, the splitting
ratio of the asymmetric beam splitter was 90/10. The laser source used in
the experiment is a commercially available OPO (OPAL from Spectra Physics) pumped by
a mode locked Ti:Sapphire laser (Tsunami from Spectra Physics). It produces
pulses of 100fs at a center wavelength of 1530nm and a repetition rate of 82MHz.
Squeezing was produced at an output pulse energy of about 27pJ for each
polarization.

\begin{figure*}
\includegraphics[scale=0.7]{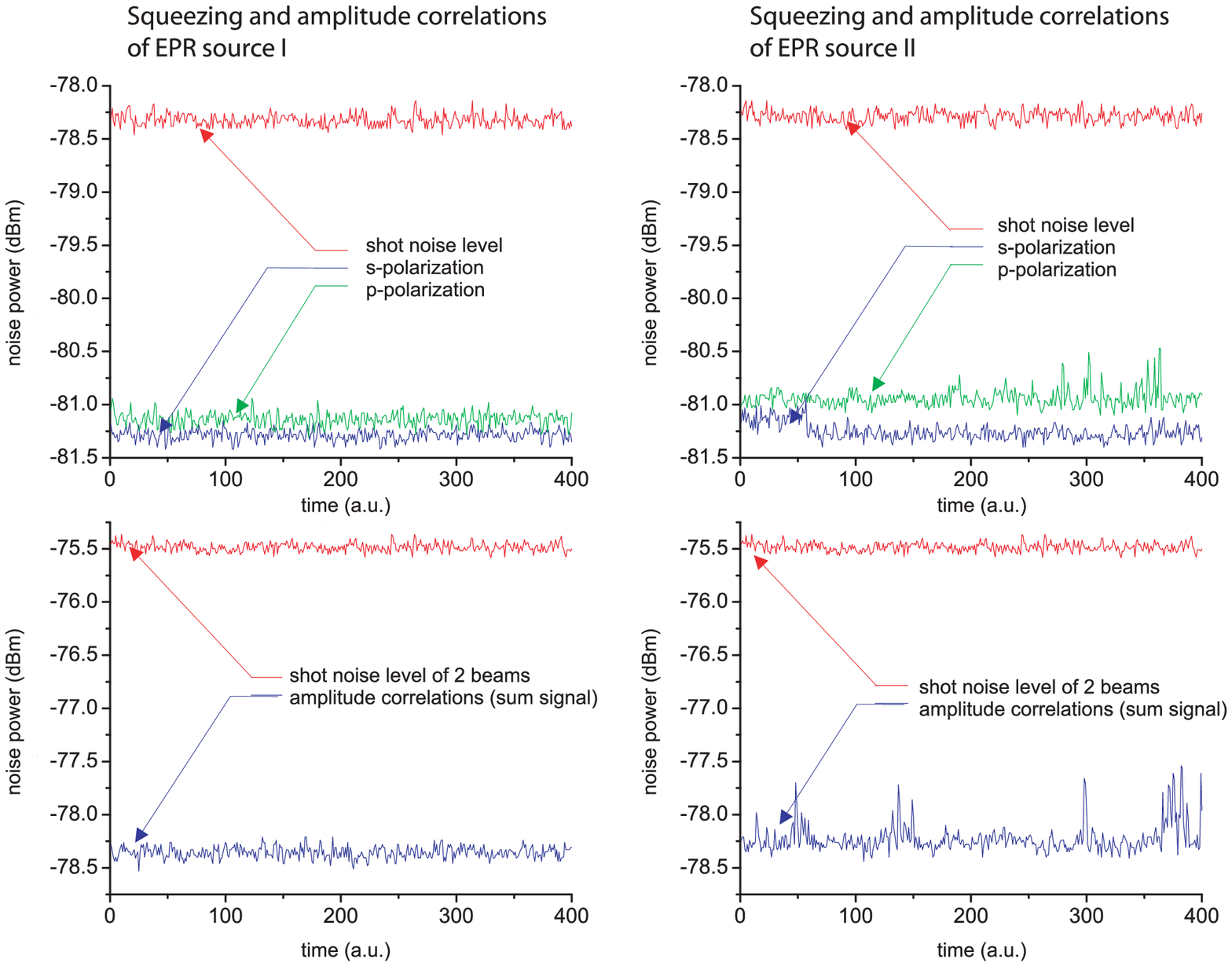}
\caption{\label{beidenolms}Characterization of the two entanglement sources:
The pictures on the top show the amplitude noise for s-- and p--polarization,
while the pictures at the bottom show the sum--signal of the EPR--entangled
beams. In all graphs, the corresponding shot noise level is depicted. Each
curve consists of 400 measurement points over a time period of 10sec. The
measured noise traces are not corrected for the electronic noise which was at
$-87.8{\rm dBm}$.}
\end{figure*}

Each squeezer produced more than 3dB amplitude squeezed light in each
polarization. This results in squeezing variances of 3dB for both entanglement
sources (see Fig.~\ref{beidenolms}). The degree of correlations was measured
for the amplitude quadrature, but due to the symmetry of the generation process,
the squeezing variances $V_{\rm sq}^+(\delta \hat{X})$ and $V_{\rm
sq}^-(\delta \hat{Y})$ for the amplitude and phase quadrature should be the
same. The peaks in the noise measurement of the entanglement source II are due
to instabilities of the laser power. The corresponding shot noise levels were
obtained by measuring the noise power when the detectors were illuminated with
coherent light of the same classical field amplitude as that of the squeezed
states. The shot noise was calibrated by illuminating the detectors with
coherent light. This calibration was checked before the experiment by comparing
the measured noise level of a coherent beam with the shot noise level determined
by the difference signal of a balanced detection setup. In the following, none
of the measured noise traces was corrected for the electronic noise which was at
$-87.8{\rm dBm}$.

The next step is the interference between the two independently generated
EPR--beams 2 and 3. The interference is a critical part of the experiment, since
a good spectral, temporal and spatial overlap of the interfering modes is
required. Therefore it is important that the beams have propagated through the
same fiber length. Moreover, both squeezers were prepared with fiber pieces from
the same coil. Still, the fiber pieces are not identical, therefore the pulses
from the different squeezers acquire different phase noise, which might be due
to different contributions from guided acoustic wave Brillouin scattering
(GAWBS) in different fiber pieces. Nevertheless it was possible to achieve a
visibility of up to 85\% and to stabilize the interference phase between EPR2
and EPR3, so that the two output modes are equally intense. This is worth
noting, as we have achieved interference between two modes of independently
generated entangled systems.

Four photodetectors were placed in the four output modes of the setup,
i.\,e.\, EPR1, EPR4, Mode5, and Mode6. Each of the detectors used a high
efficiency InGaAs--photodiode (Epitaxx ETX 500). To avoid saturation of
the detectors, a Chebycheff lowpass filter was used with a cutoff frequency of
about 35MHz to suppress the high noise peaks at the repetition frequency and the
harmonics of the laser. The photocurrent fluctuations were detected at
a frequency of 17.5MHz with a resolution bandwith of 300kHz. The
measurements were averaged with a video bandwith of 30Hz, the measurement time
for each trace was 10s.

In the experiment, we examined the correlations between the two output modes
EPR1 and EPR4. The various noise levels are compared with the corresponding shot
noise levels and the noise levels of the individual beams. First, the noise
level of the sum signal of beams EPR1 and EPR4 was measured and compared with
the noise power of the individual beams, i.\,e.\,the variances of $i1$ and $i4$
and $i1+i4$ were measured, not yet taking into account the result from the
Bell--state measurement, $i_{\rm Bell}^{+}$. This corresponds to the case of the
full entanglement swapping protocol with respect to the amplitude
quadrature only without transmission of the classical information from the
Bell--state measurement.

\begin{figure}
\includegraphics[scale=0.65]{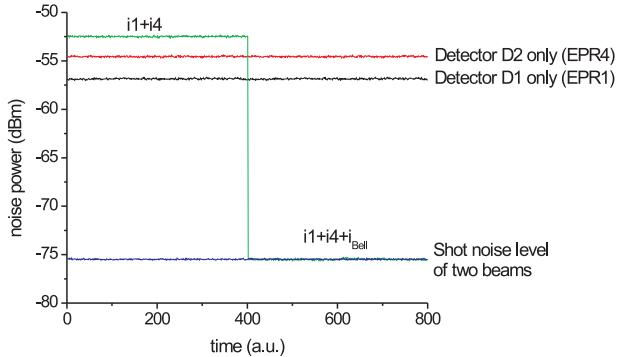}
\caption{\label{entswapmess}Noise power of different combinations of the
output modes: Traces of the single beams EPR1 and EPR4 are shown together with
the combined shot noise level of two beams. On the left side, the sum signal
of EPR1 and EPR4 is detected, on the right side, the signal from the
Bell--measurement is added to the photocurrent detected on mode EPR4. The signal
trace incidently coincides with the shot noise level of two beams. Each part of
the trace (left and right) consists of 400 measurement points over a time period
of 10s. The measured noise traces are not corrected for the electronic noise
which was at $-87.8{\rm dBm}$.}
\end{figure}

There are no correlations between the two modes EPR1 and EPR4, i.\,e.\,
\begin{equation}
V(\hat{X}_{\rm EPR1}+\hat{X}_{\rm EPR4})=V(\hat{X}_{\rm
EPR1})+V(\hat{X}_{\rm EPR4})
\end{equation}
which can be seen on the left side of Fig.~\ref{entswapmess}, and which is
obvious as the two entanglement sources are independent.

The noise power of the beams EPR1 and EPR2 shows that each single beam is very
noisy. This indicates that the squeezer does not generate minimum uncertainty
states, but states which acquire excess phase noise of about
20dB, an effect that originates from the propagation of the light through the
fiber. Although we tried to set up an exact replica of the first squeezer, the
noise power of the generated EPR--beams of the second squeezer is different
from that of the first squeezer.

In a second step, the correlations between EPR1 and EPR4 are checked again, this
time taking into account the result of the Bell--measurement. The
photocurrent of EPR4 is modified to give $i4+i_{\rm Bell}^{+}$. This
photocurrent is added to $i1$. The noise power of the resulting photocurrent
drops to the shot noise level of two beams, indicating that there are strong
correlations between $i1$ and $i4+i_{\rm Bell}^{+}$. The noise levels are
plotted on the right side of Fig.~\ref{entswapmess}.
This situation corresponds to the case where the classical information gained
from the Bell--state measurement is transmitted and mode EPR4 is
modulated and the correlations between OUT1 and OUT2 are measured (compare
Fig.~\ref{setup}). This result clearly indicates that the full entanglement
swapping experiment should be possible. With the given squeezing values, a
squeezing variance $V_{\rm sq}^{+}(\delta X)=1$ is expected for the
output modes. The limitation in the experiment so far is the
relatively low detected input squeezing of 3dB, otherwise correlations below the
shot noise level would be visible. However, there is the 3dB penalty due to the
fact that the squeezed states used have a high degree of excess phase noise. The
relatively low resulting squeezing values are partly caused by the
detection scheme, where at least three detectors were involved in the
measurement of each individual noise trace, which causes some balancing problems
of the electronics.

Although the squeezing variance expected in the full experiment is at
the shot noise limit, there are hints that the teleportation
of a highly non--classical state with continuous variables is possible.
We can speak of successful quantum teleportation of beam EPR2 on beam EPR4 in
the following sense: Let us compare the measurement results with those that can
be obtained by classical teleportation, where the amplitude and phase
fluctuations of the mode to be teleported are measured simultaneously and an
independent coherent beam is modulated. Squeezing variances between OUT1 and
OUT2 which are 1.77dB higher than those that were measured in our experiment are
expected with the given degree of correlations of 3dB. This can be understood by
a similar argumentation to that given by Furusawa et al.\,\cite{FUR98}. In
classical teleportation, two extra units of vacuum are mixed into the system,
one from the Bell--measurement and one from the coherent beam that is modulated.

\subsection{Interpretation of the results as multipartite correlations}

Another aspect of the experiment is that the measurement results can
also be interpreted in a different way. The setup is as before, two pairs of
entangled beams are generated and the beams EPR2 and EPR3 are combined at a
50/50 beam splitter. We are now interested in the correlations of the complete
four--mode state.

\begin{figure}
\includegraphics[scale=0.4]{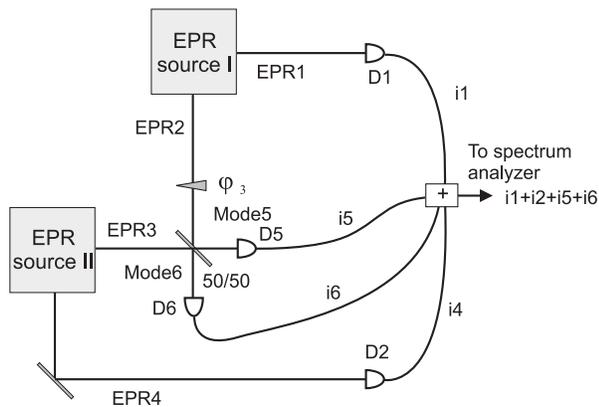}
\caption{\label{4fachkorr}Generation of a highly correlated 4--partite state}
\end{figure}
\begin{figure}
\includegraphics[scale=0.65]{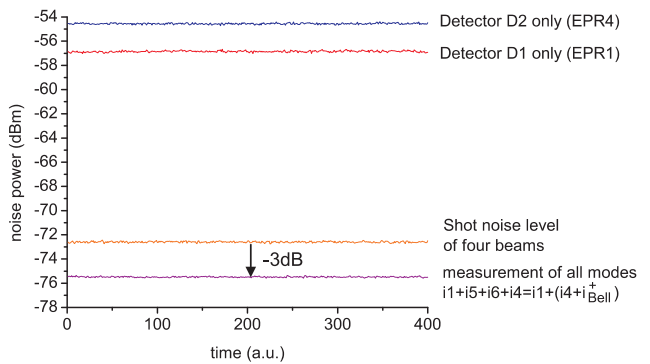}
\caption{\label{4fachkorrmess}Four--partite correlations: The measurement of
all four modes gives a signal that is 3dB below the corresponding shot noise
level. Thus the generated four--mode state is highly correlated, each
individual beam having a high noise level. The measurement time was 10s. The
measured noise traces are not corrected for the electronic noise which was at
$-87.8{\rm dBm}$.}
\end{figure}

To characterize the correlations, a detector is placed in each output mode, that
is in EPR1, EPR4, Mode5 and Mode6, and the photocurrents are added as depicted
in Fig.~\ref{4fachkorr}. The situation is completely equivalent to the case
where the Bell--measurement was performed and the photocurrents were added.
However, the sum--photocurrent of all modes shows that the four mode state has a
high degree of correlations, because the sum signal shown in
Fig.~\ref{4fachkorrmess} is 3dB below the corresponding shot noise level of four
equally bright coherent states:

\begin{equation}\label{corr}
V(\hat{X}_{\rm
EPR1}+\hat{X}_{\rm Mode5}+\hat{X}_{\rm Mode6}+\hat{X}_{\rm
EPR4})<4V(\hat{X}_{\rm coh}).
\end{equation}
This indicates that a highly correlated four--mode state was created. The
question that arises now is whether this state is genuinely four--partiy
entangled.

A genuinely multipartite entangled state means that none of the parties is
separable from any other party in the total density operator. For example, the
two initial EPR sources in our setup correspond to an entangled four--mode state
which is not genuinely four--party entangled. Though none of the modes is
completely separable from the rest, the total state only consists of two
two--party entangled states, mathematically desribed by the tensor product of
two bright EPR states. In contrast, the output four--mode state of modes EPR1,
Mode5, Mode6, and EPR4, is, in principle, a genuinely four--party entangled
state, at least when assuming pure input states. This can be understood most
easily by examining the corresponding four--mode Wigner function
$W(\alpha_1,\alpha_5,\alpha_6,\alpha_4)$ of the output state. Assuming two pure
initial EPR sources, the output state $W(\alpha_1,\alpha_5,\alpha_6,\alpha_4)$
is also pure. For pure states, the non--factorizability
of the Wigner function,
$W(\alpha_1,\alpha_5,\alpha_6,\alpha_4)\neq W(\alpha_1)
W(\alpha_5,\alpha_6,\alpha_4)$ and
$W(\alpha_1,\alpha_5,\alpha_6,\alpha_4)\neq W(\alpha_1,\alpha_5)
W(\alpha_6,\alpha_4)$ for all permutations of $(1,5,6,4)$,
proves the genuine four--party entanglement.
The output Wigner function for the special case
of two two--mode squeezed vacuum states as the input
EPR--sources is given in Ref.~\cite{LOO02a}.
For any degree of the squeezing,
it cannot be written in any product form hence the state is genuinely
four--party entangled. Moreover, partial transposition
(partial sign change of the momentum quadratures \cite{SIM00})
applied to its four--mode correlation matrix with respect
to any possible splitting of the four modes \cite{LOO02,GIE01b}
always leads to an unphysical state, thus ruling out
any partial separability.

The four--party correlations of the four--mode squeezed vacuum, as can be seen
in the Wigner function of Ref.~\cite{LOO02a}, lead to quadrature
triple correlations such as $V[\hat X_1-(\hat X_5 + \hat
X_6)/\sqrt{2}]\rightarrow 0$ and
$V[(\hat X_5 - \hat X_6)/\sqrt{2} - \hat X_4]\rightarrow 0$.
In our experiment, due to the brightness of the beams and the
direct detection scheme employed,
the four--party correlations become manifest
in a combination of the quadrature amplitudes of all four modes,
as described by equation (\ref{corr}).
Whether the mixedness of our entangled four--mode state due to the
excess phase noise causes a significant deterioration
of the four--party entanglement should be further investigated,
also with respect to the experimental verification
of the four--party entanglement (see below). Since the genuine four--party
entanglement is present for any non--zero squeezing in the pure four--mode
squeezed vacuum state \cite{LOO02}, we expect that the mixed state in our
experiment remains genuinely four party entangled.

\begin{figure}
\includegraphics[scale=0.65]{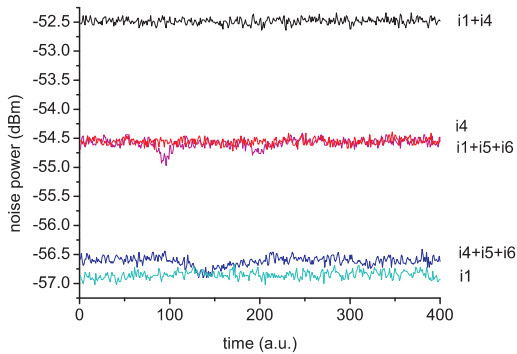}
\caption{\label{dreistrahlen}Correlations of the four--mode state.
Measurement of three modes compared with the noise levels of the indivdual
beams. The measurement time was 10s. The shot noise level for three modes is
about $-73.7{\rm dBm}$.}
\end{figure}

To further characterize and describe the correlations of the four--mode state,
we measured the noise of the combination of three beams, that is the variance of
the photocurrents $i1+i5+i6=i1+i_{\rm Bell}^{+}$ and $i5+i6+i4=i_{\rm
Bell}^{+}+i4$, which are plotted in Fig.~\ref{dreistrahlen}. These signals
exhibit a high noise level compared with the shot noise level. To be more
precise, the noise level of $i1+i5+i6$ equals the noise level of $i4$ (EPR4) and
$i5+i6+i4$ equals $i1$ (EPR1). This behaviour becomes clear by the following
considerations: The sum--signal from Mode5 and Mode6, i.\,e.\, $i_{\rm
Bell}^{+}$, contains the full amplitude fluctuations of EPR2 and EPR3. Thus by
taking the sum with EPR1, only the fluctuations originating from EPR3 remain, as
EPR1 and EPR2 were initially entangled and the fluctuations cancel almost
completely compared with the noise of a single beam. Thus the measured noise
level of $i1+i5+i6$ is equal to the fluctuations of beam EPR3 which has the same
noise level as beam EPR4. Similarly, the noise level of the combined measurement
of $i5+i6+i4$ can be explained. This measurement is consistent with the fact
that the excess noise of the individual EPR--beams of the different EPR--sources
is slightly different. As mentioned before, there are no correlations between
the beams EPR1 and EPR4, as the variance of $i1+i4$ is just the sum of the
variances of $i1$ and $i4$.

\section{Discussion and outlook}

\subsection{Phase measurement: Full proof of entanglement swapping }
The full proof of entanglement swapping will be more challenging as not only
the correlations in the amplitude quadrature, but also those in the
phase quadrature must be checked. For this purpose, a real modulation of beam
EPR4 is needed. As intense entangled beams are used, no local oscillator can be
employed to measure correlations in the phase quadrature due to the saturation
of the detectors. Instead, some interferometric scheme is needed to measure the
correlations in the phase quadrature. This can be achieved by letting the two
output modes, OUT1 and OUT2, interfere at another 50/50 beam splitter.

\begin{figure*}
\includegraphics[scale=0.4]{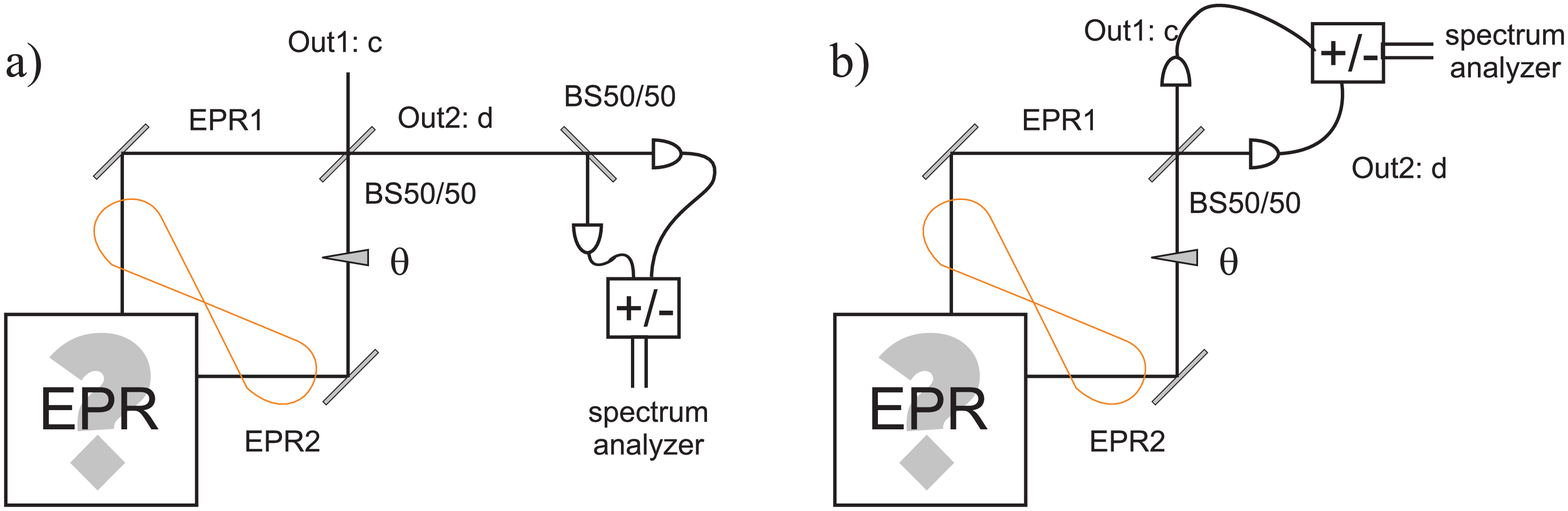}
\caption{\label{phase} Two possible setups for an indirect measurement of
correlations in the phase quadrature of two beams EPR1 and EPR2. Both input
beams are superimposed on a 50/50 beam splitter. In a) the noise variance of one
output port is detected. This also provides a direct measure for the
non--separability of states EPR1 and EPR2. In b) the signal from both output
ports is detected and the difference channel provides information about the
correlations of the phase quadrature of the beams EPR1 and EPR2.}
\end{figure*}

Two possible interferometric detection schemes will be described to test
whether a pair of beams, here called EPR1 and EPR2, coming from a black box are
entangled or not. In the first case, the noise power can be measured in one of
the output ports of the interfero\-meter using balanced detection
(compare Fig.~\ref{phase}a). When the correlations in the amplitude
quadratures of EPR1 and EPR2 have already been measured, correlations in the
phase quadrature can be determined from an amplitude quadrature measurement of
OUT1 or OUT2. If squeezing occurs, then also the phase quadrature is correlated
and the beams EPR1 and EPR2 must be entangled \cite{SIL01}. It is also possible
to check directly if the outgoing beams are non--separable: The variance of the
sum photocurrent measured at the output of the interferometer is proportional to
the sum of the squeezing variances, see Eq.(\ref{peres}). Thus the Duan
non--separability criterion can be verified in a single
measurement\cite{KOR02a}. It was shown that, if squeezing occurs for a certain
interference phase $\theta$, the states EPR1 and EPR2 are necessarily entangled
\cite{KOR02a}, a consequence of the non--separability criterion for continuous
variables.

The other possibility is to put a detector in each output port and record the
sum and the difference of the photocurrents (see Fig.~\ref{phase}b). The
difference signal is proportional to the variance $V(\hat{Y}_{\rm EPR1}
-\hat{Y}_{\rm EPR2})$ which drops below the shot noise level if there are
non--classical correlations in the phase quadrature. The correlations in the
amplitude quadrature can be checked simultaneously by taking the sum of the
photocurrents. This measurement is of the same type as the Bell--state
measurement in the entanglement swapping process. This detection scheme is also
used for dense coding experiments \cite{LI02} and was proposed for a
continuous variable entanglement swapping experiment \cite{ZHA02a}.

\begin{figure}
\includegraphics[scale=0.4]{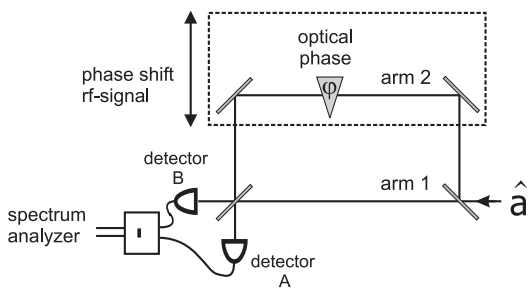}
\caption{\label{phase4} Unbalanced interferometer for detection of phase
fluctuation without local oscillator.}
\end{figure}

Alternatively, phase detection $\delta \hat{Y}_{\rm OUT1}$ and $\delta
\hat{Y}_{\rm OUT2}$ can be performed on beams OUT1 and OUT2 respectively. For
this purpose, an extremely unbalanced interferometer can be used \cite{TRA98},
which maps phase fluctuations onto amplitude fluctuations for certain
measurement frequencies. At the repetion rate of $82$MHz of our laser system, an
arm length difference of about $7.2$m is required to measure phase fluctuations
at a radio frequency of $20.5$MHz. In that case, a phase shift is accumulated
which leads to a 90$^{\circ}$ rotation of the quantum uncertainty sideband.
After interference and together with the detection scheme, the phase noise of
the input beam can be recorded in direct detection (see Fig.~\ref{phase4}). Thus
the phase shift for the rf--signal has to be adjusted as well as the optical
interference phase such that the two output ports are equally intense. With such
an interferometer, it will be possible to avoid phase modulation, as in the case
of the amplitude fluctuations, and check directly the obtained photocurrents
from the phase measurement together with the photocurrent from the
Bell--measurement $i_{\rm Bell}^-$ for nonclassical correlations.

\subsection{Criteria for multipartite entanglement}

The four--party correlated state generated by the interference
of two two--party entanglement sources could be an important
requisite for multipartite quantum communication and networking.
However, it should be first verified experimentally that the
generated state is indeed genuinely four--party entangled.
For Gaussian states, the full characterization of multipartite
entangled states requires the detection of all independent
entries of the correlation matrix. However, in Ref.~\cite{LOO02},
inequalities are derived which impose necessary conditions
onto a multi--party multi--mode state to be partially or fully separable.
Hence violations of these conditions rule out any form
of separability, thus being sufficient for genuine
multipartite entanglement. These criteria involve measurements
of linear combinations for the two conjugate quadratures
of all modes. The conditions are expressed in terms of the sum
of the variances of these combinations.
This is similar to the Duan criterion stated in equation (\ref{peres}),
where the sum of the squeezing variances of the $\hat X$ and
$\hat Y$ quadratures is considered.
Such criteria are only sufficient and not necessary for entanglement,
but they apply to arbitrary quantum states (pure, mixed,
Gaussian, or non--Gaussian).
By using criteria based on quadrature linear combinations
for the verification of genuine multipartite entanglement,
the measurement of the entire correlation matrix of a Gaussian
state is no longer needed.
However, in general, these criteria cannot be readily
applied to direct detection schemes.
Hence simple experimental criteria and measurement techniques
like those described above for the two--mode case still
have to be developed for the multi--mode case.

\section{Conclusions}
To summarize, we have adapted a scheme to perform entanglement swapping with
intense, pulsed, quadrature entangled beams using direct detection. By
substituting the required optical modulation with an adapted detection setup we
have shown experimentally that the amplitude correlations behave as they
should for entanglement swapping. This is a strong hint for
entanglement between the two output states, which have never interacted
directly. To fully prove entanglement, the phase correlation still
has to be measured. The joint analysis of the
four individual modes shows that we have generated and partly characterized a
novel four partite continuous--variable state that exhibits four--fold
correlations below shot noise in the amplitude quadrature. Although so far
there is no experimental proof that this state is really four--party entangled,
the measured correlations support this conjecture. We have taken an experimental
step towards entanglement swapping with continuous variables and produced an
intense light state showing four--fold quantum correlations which is consistent
with genuine four--party entanglement.

\begin{acknowledgments}
This work was supported by the Schwerpunkt Programm 1078 of the Deutsche
Forschungsgemeinschaft and by the EU grant under QIPC, project
IST-1999-13071 (QUICOV). The authors thank M. Langer and T. Lang for the help
with the detector electronics and J. Trautner for useful discussions about the
phase measuring interferometer. Qing Pan gratefully acknowledges financial
support from China Scholarship Council.
\end{acknowledgments}

\newpage 

\end{document}